\documentclass[draft]{agujournal2019}
\usepackage{url} 
\usepackage{lineno}
\usepackage[inline]{trackchanges} 
\usepackage{soul}
\draftfalse

\journalname{be submitted soon}

\begin{document}

%
%

\title{First results of Low-latitude Ionospheric Irregularities measured by NavIC and GPS near the Anomaly Crest and the Magnetic Equator} 

%
%





\authors{Sumanjit Chakraborty\affil{1,2}, Abhirup Datta\affil{1}, Deepthi Ayyagari\affil{1}}

\affiliation{1}{Department of Astronomy, Astrophysics and Space Engineering, Indian Institute of Technology Indore, Simrol, Indore 453552, Madhya Pradesh, India}

\affiliation{2}{Space and Atmospheric Sciences Division, Physical Research Laboratory, Ahmedabad, 380009 Gujarat, India}




\correspondingauthor{Sumanjit Chakraborty}{sumanjit11@gmail.com}




\begin{keypoints}
\item Low-latitude ionospheric irregularity studies are performed using NavIC and GPS signals for the first time.
\item Measurements of disturbed and quiet-time low-latitude ionospheric irregularity scale sizes using actual transionospheric signals are presented.
\item No correlation found between the scintillation characteristics derived from the data and the geomagnetic activity. 
\end{keypoints}

%
%

%
%


\begin{abstract}

Ionospheric irregularity studies are important aspects for understanding ionospheric physics and related processes, especially near the low-latitude regions. However, simultaneous measurements (utilizing the L-band signals of NavIC and GPS) of irregularity scale sizes over the Indian longitude sector, has not been addressed extensively. To address this problem, the paper presents simultaneous characterization of low-latitude ionospheric irregularities over a location near the northern EIA crest (Indore: 22.52$^\circ$N, 75.92$^\circ$E geographic and magnetic dip of 32.23$^\circ$N) and a location (Hyderabad: 17.41$^\circ$N, 78.55$^\circ$E geographic and magnetic dip of 21.69$^\circ$N) between the crest and the magnetic equator, utilizing the Indian navigation system, NavIC and GPS L5 signal C/N$_o$ variations to determine the range of the ionospheric irregularity scale sizes using Power Spectral Density (PSD) analysis. The study period spans from September 2017- September 2019, covering both disturbed and quiet-time conditions in the declining phase of solar cycle 24. Observations show that the irregularity scale size ranges from about 500 m to 6 km. Furthermore, no correlation was observed between the data derived scintillation characteristics and the geomagnetic activity. This study for the first time, shows the nature of the temporal PSD for ionospheric scintillation during varying solar and geophysical conditions, by measuring the irregularity scale sizes utilizing simultaneous observations from NavIC and GPS from locations near the northern crest of the EIA and in between crest and the magnetic equator, ensuring proper characterization of ionosphere over the geosensitive Indian subcontinent.
\end{abstract}


%
%

%


%
%
%
%


\section{Introduction}

The phenomenon of scintillation can be thought analogue with the stars that twinkle in the night sky due as a result of variations in density of the atmosphere due to turbulence. 
The phenomenon of plasma instability (post-sunset) in the equatorial ionosphere generates irregularities and large scale depletions in the electron density referred to as the Equatorial Plasma Bubbles (EPBs) (\citeA{sc:0} and references therein). Radio waves propagating through these irregularities experience diffraction and scattering, which cause random fluctuations in the VHF and L-band signal amplitude, phase, direction of propagation, and polarization referred to as scintillations \cite{sc:3,sc:4,sc:8}. It is well known that the irregularities which produce scintillations are mainly found in the F layer of the ionosphere ($\sim$ 300-350 km), where the plasma density has the maximum value. 
Rayleigh–Taylor (R-T) instabilities are considered as the primary mechanism generating EPBs with scale sizes from a few hundreds of meters to several kilometers \cite{sc:8}.
Activities of scintillations are generally observed during the period of maximum solar activity, occurring near the magnetic equator in the post-sunset-to-midnight sector \cite{sc:6,sc:7}. 
It is well known that ionospheric scintillations are demonstrations of space weather effects, affecting the performance of space-based navigation and communication systems which rely on transionospheric radio-wave propagation \cite{sc:4,sc:5}. 

For characterizing radio wave scintillation that travels in the ionosphere, numerous phase screen theories have been developed, as early as the 1950s following works done by \cite{sc:1,sc:2}. 
Generally, scintillation of radio waves is calculated using the theory of wave propagation in random media to obtain the parameters of the exiting wave from the ionosphere. It is performed by solving the Fresnel diffraction theory problem involving the propagation of these waves traveling between the ionosphere and the Earth. Thus, these calculations replace the ionosphere by an equivalent phase screen that is random, where the irregularities in the ionosphere are considered as phase objects \cite{sc:10}. Simultaneously, the ground pattern produced by radio waves propagating through this screen is derived from the diffraction theory \cite{sc:11}.

In general, the one-and two-dimensional phase screens are analytic representations of the intensity of the spectral density function in terms of the various phase structure function combinations \cite{sc:9}. Phase screen models have been extensively used to simulate the distortion of wideband waveforms for communications \cite{sc:26}, fading of GNSS satellite signals, scintillation of satellite signals used in radio occultation experiments, to name a few. These phase screen models' inputs are the in-situ electron density measurements, a radio receiver's time series phase measurements, and a turbulent ionospheric medium-based stochastic model. Several researchers (\citeA{sc:3,sc:4,sc:5,sc:29} and references therein) have addressed in-situ electron density fluctuations measured by the satellites to predict the propagation effects. However, the disadvantage lies in the fact that measured values sample the electron density irregularities and fluctuations at the satellite's orbital altitude. The transionospheric propagation effects can be attributed to the integrated development of the density variation along the signal ray path at all altitudes. If any, the relation between density fluctuations sampled at one altitude to another is still not evident in the literature. Therefore, if the density irregularities associated with an EPB have not risen to the satellite's altitude that measures the in-situ density values, detection of fluctuations will be absent despite the EPB causing radio scintillations as a result of the irregularities present at lower altitudes.
There is an advantage of probing the ionosphere at all altitudes, from the ground-based observations of signal phase transmitted by a satellite at the topside ionosphere or beyond, over in-situ electron density observations. Thus, it becomes essential to understand and quantify the contribution from phase scintillations caused by diffraction, in addition to interpreting the signal phase time series and the corresponding limitations of judiciously representing the structure of irregularity at these altitudes \cite{sc:25,sc:4,sc:21,sc:20,sc:23,sc:24}.

The presence of the northern crest of the Equatorial Ionization Anomaly (EIA) and the geomagnetic equator that touches the southern tip of the Indian peninsular region, accompanied by sharp latitudinal gradients of ionization, makes the Indian longitude sector a highly geosensitive region of investigation for ionospheric research during geomagnetically disturbed periods when the low-latitude ionization is significantly affected as a result of solar eruptions like the Coronal Mass Ejections (CMEs), the Coronal Hole/High-Speed Solar Wind streams (CH/HSSW) and associated Corotating Interaction Regions (CIRs) \cite{sc:12,sc:13,sc:19}. Previous studies show the development of power law phase screen theories and models for scintillation based on weak and strong conditions. The nature of phase screen, associated with both weak and strong scattering, in the geosensitive region of India, during geomagnetically disturbed and quiet conditions, has not been extensively studied in the literature. This study for the first time, to the best of our knowledge, shows the nature of the temporal Power Spectral Density (PSD) for ionospheric scintillation associated with weak as well as strong scatter conditions, during varying solar and geophysical conditions, by measuring the irregularity scale sizes utilizing the Navigation with Indian Constellation (NavIC) as well as the Global Positioning System (GPS) observations from locations near the northern crest of the EIA and in between crest and the magnetic equator that ensure proper simultaneous ionospheric irregularity characterization over the geosensitive Indian longitude sector.

\clearpage
\section{Data Collection and Analysis Procedure}

In this work, we started by investigating the ionospheric activity above Indore within the period September 01, 2017 through September 30, 2019, which coincides with the declining phase of solar cycle 24. 
Data has been collected using a multi-constellation and multi-frequency GNSS receiver (Septentrio PolaRxS Pro) capable of receiving GPS L1 (1575.42 MHz), L2 (1227.60 MHz) and L5 (1176.45 MHz) signals and specialized for ionospheric monitoring and space weather research. Here, the amplitude and phase samples get generated at 50 Hz in order to calculate the scintillation indices, C/N$_o$ (dB-Hz), satellite lock time, azimuth (deg), elevation (deg), and the Total Electron Content (TEC). We followed the similar procedure as done by \cite{sc:28}
We have also taken data using a NavIC (ACCORD) receiver, capable of receiving L5 and S1 (2492.028 MHz) signals along with GPS L1 signal. The output of the receiver includes the C/N$_o$, azimuth, elevation, pseudorange (in m), carrier cycles (cycles). The receiver is provided by the Space Applications Centre, ISRO. 
Both these receivers are operational in the Department of Astronomy, Astrophysics and Space Engineering of Indian Institute of Technology Indore (22.52$^\circ$N, 75.92$^\circ$E geographic; magnetic dip 32.23$^\circ$N). Additionally, data from another NavIC receiver operational at the department of Electronics and Communication Engineering, Osmania University, Hyderabad (17.41$^\circ$N, 78.55$^\circ$E geographic; magnetic dip 21.69$^\circ$N), has been used for the analysis. 
From the entire analysis of the period of 25 months, we have only found scintillation events on 27 days as observed from NavIC Indore and validated by GPS Indore and NavIC Hyderabad. In the current work, we have used the 27 scintillation events as our parent sample. Among those 27 events, we have chosen three representative events (with most coverage from all the GPS and NavIC satellites). In the rest of the paper, we will show results and further analysis of these three events which are suitably chosen across different solar and geophysical conditions.

\section{Results and Discussions}

In this section, the strong storm day of September 8, 2017, the weak storm day of August 17, 2018 and the quiet day of September 16, 2017 have been presented as sample cases to study the C/N$_o$ variation as observed by the NavIC satellites; thereafter analysing the corresponding PSD and the respective irregularity scale sizes with the spectral slopes, under varying geophysical conditions over Indore: a location near the northern crest of the EIA of the Indian longitude sector. It is to be noted that as the NavIC satellites are geostationary/geosynchronous, the C/N$_o$ data can be continuously observed throughout the day. Similar approaches are taken to show the respective variations as observed by the NavIC data over Hyderabad: a location away from the northern crest, for the September 8, 2017 storm day. Additionally for comparison of NavIC variation to that of the GPS, same methodology is adopted for the GPS observations on these days over Indore.  

\subsection{September 8, 2017: a case of strong storm}

Due to a CME that arrived on September 6, 2017, a G4 (K$_p$= 8, severe) level geomagnetic storm was observed at 23:50 UT on September 7, 2017, at 01:51 UT and 13:04 UT on September 8, 2017, according to the National Oceanic and Atmospheric Administration (NOAA) (https://www.swpc.noaa.gov/). The Disturbance storm time (Dst) index (available at http://wdc.kugi.kyoto-u.ac.jp/dstdir/), which measures the strength of the disturbance caused to the horizontal component of the geomagnetic field due to incoming solar eruptions, showed dips during September 8, 2017, with values going below -100 nT (-124 nT at 02:00 UT and -109 nT at 18:00 UT). This signified that the nature of the geomagnetic disturbance had been strong in severity and double-peaked.

\begin{figure}[ht]
\noindent\includegraphics[width=5in,height=3.5in]{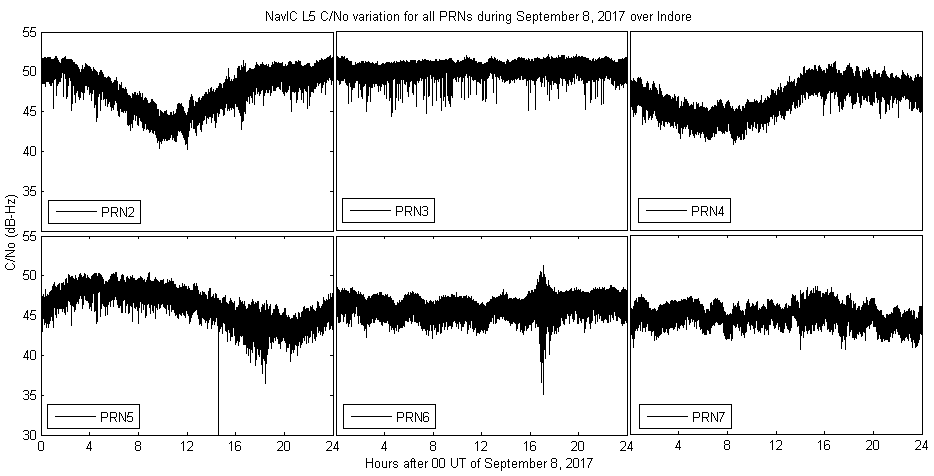}
\caption{The C/N$_o$ (dB-Hz) variation during the disturbed day of September 8, 2017, as observed by L5 signal of NavIC satellite PRNs 2-7 over Indore. It is to be noted here that LT=UT+05:30h.} 
\label{sc1}
\end{figure}

Figure \ref{sc1} shows the C/N$_o$ variation (dB-Hz) over Indore located near the northern crest of the EIA that passes over the highly geosensitive Indian longitude sector, for the entire  day of September 8, 2017, as observed by a set of geostationary (PRNs 3,6 and 7) and geosynchronous (PRNs 2,4 and 5) satellites of NavIC. The observations are taken by utilizing the L5 (1176.45 MHz) signal of NavIC. Upon closely observing panels showing PRNs 5 and 6 of Figure \ref{sc1}, it is clearly visible that there had been strong fluctuations in the C/N$_o$ with values dropping from $\sim$52 dB-Hz to $\sim$35 dB-Hz around 17-18 UT (22:30-23:30 LT), between local post-sunset and pre-midnight sector. 

\begin{figure}[ht]
\centering
\noindent\includegraphics[width=5.5in,height=4in]{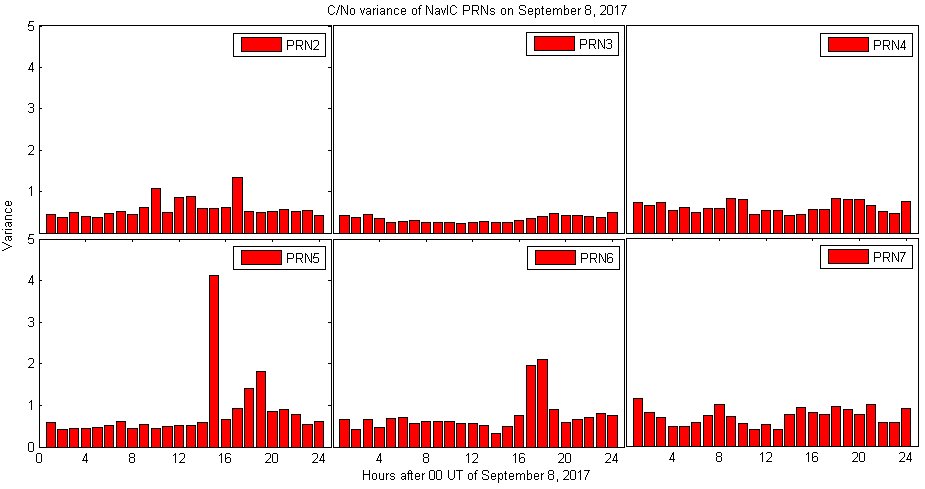}
\noindent\includegraphics[width=5.5in,height=4in]{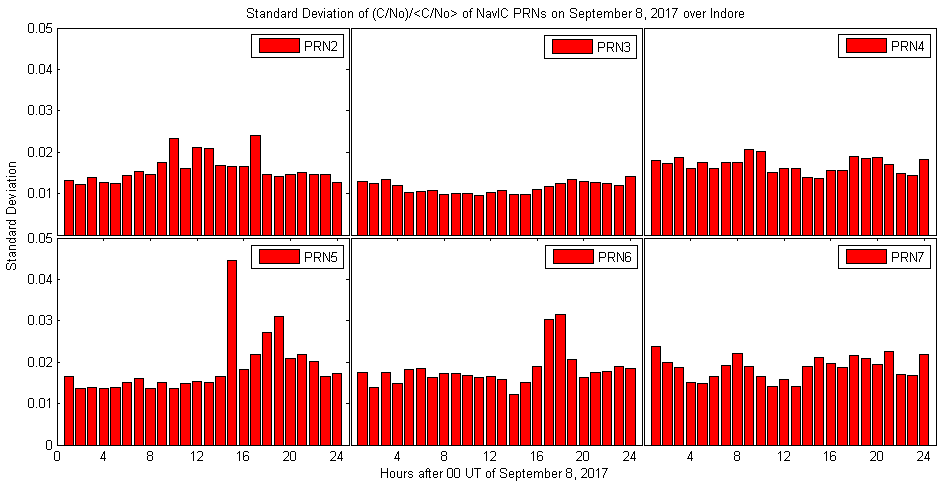}
\caption{The hourly binned variance plots (top panel) of C/N$_o$ and the corresponding standard deviation plots (bottom panel) of $\frac{(C/N_o)}{<C/N_o>}$ for all PRNs of NavIC on September 8, 2017 as observed from Indore.} 
\label{sc2}
\end{figure}

To verify whether the observed C/N$_o$ drops in Figure \ref{sc1} are due to the scintillations that had occurred during this period, top panel of Figure \ref{sc2} shows the hourly binned variance plots of all the PRNs of NavIC during the entire day of September 8, 2017, while the bottom panel of the same figure shows the corresponding standard deviation of the normalized ratio $\frac{(C/N_o)}{<C/N_o>}$ ($<>$ indicating average values). This parameter is similar to the scintillation index $S_4$. In the hourly bin of 18-19 UT for PRN 5 and 17-18 UT for PRN 6, it is clearly visible that the variance over the average values were significant. It is to be noted that the high value observed in the 14-15 UT bin for PRN5 has not been considered as it is getting manifested as a result of the C/N$_o$ drop to zero observed by this PRN in Figure \ref{sc1}.  

\begin{figure}[ht]
\centering
\noindent\includegraphics[width=3in,height=2.5in]{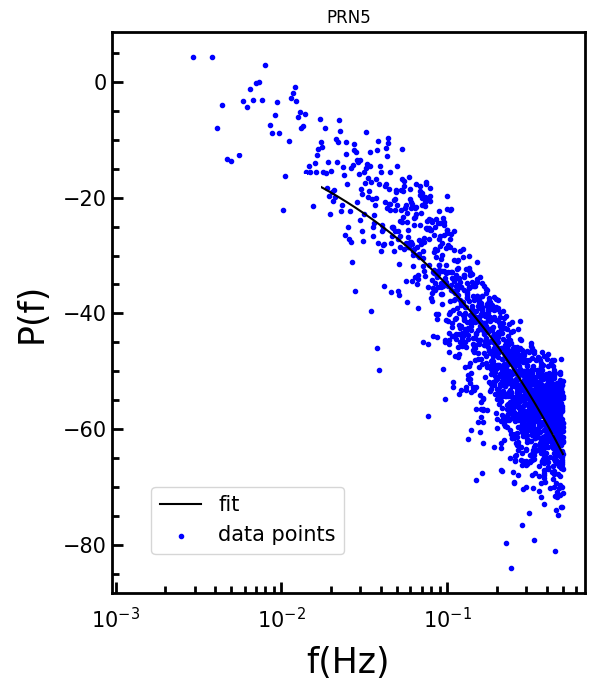}
\noindent\includegraphics[width=3in,height=2.5in]{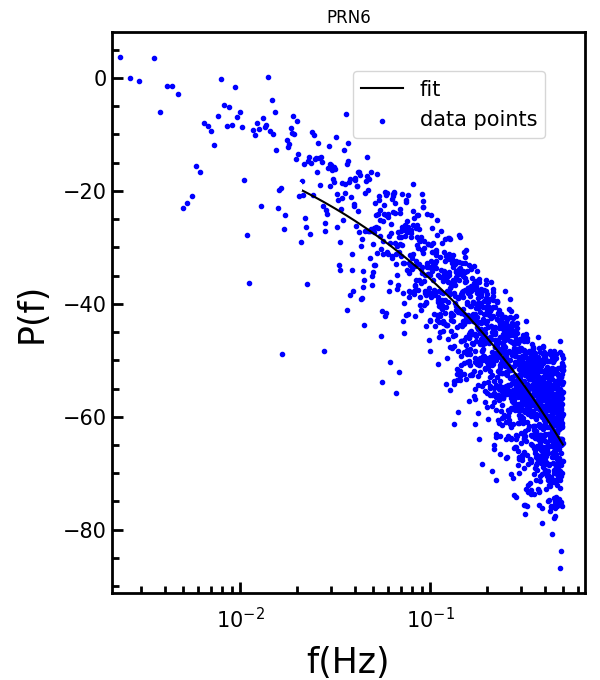}
\caption{The PSD variations with least square fit (black solid line) corresponding to the bins of intense C/N$_o$ variation of Figure \ref{sc2}.} 
\label{sc3}
\end{figure}

\begin{figure}[ht]
\centering
\noindent\includegraphics[width=5.5in,height=3.3in]{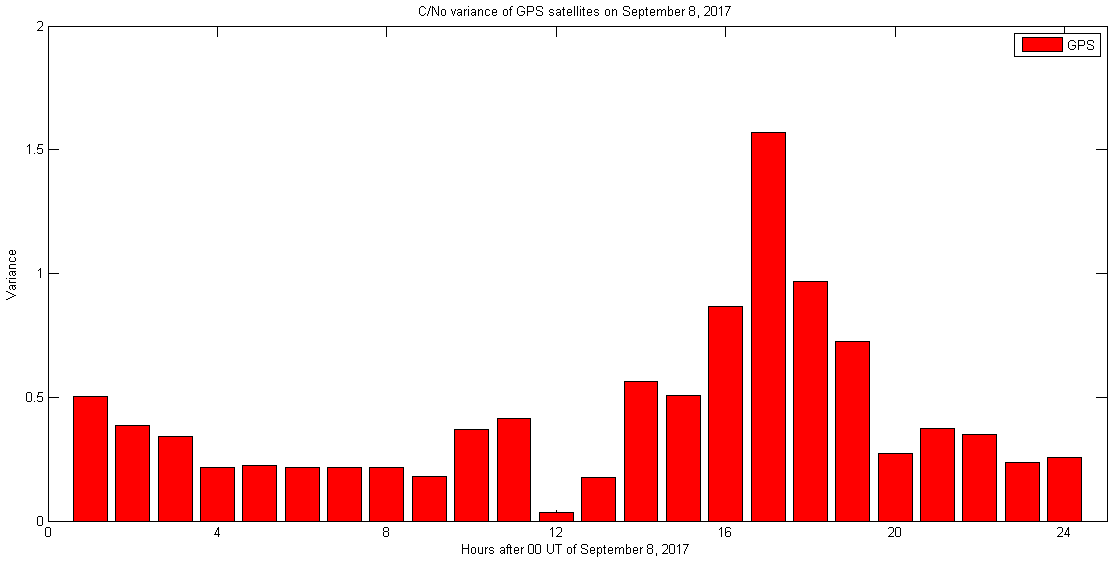}
\noindent\includegraphics[width=5.5in,height=3.3in]{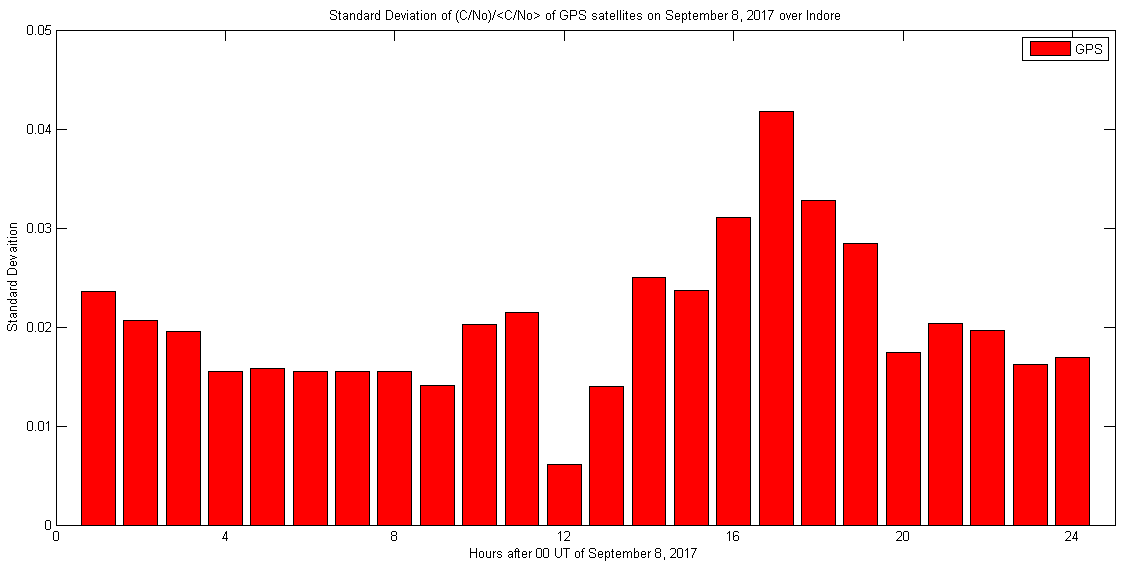}
\caption{The hourly binned variance plots (top panel) of C/N$_o$ and the corresponding standard deviation plots (bottom panel) of $\frac{(C/N_o)}{<C/N_o>}$ for the GPS satellites on September 8, 2017 over Indore.} 
\label{sc4-gps}
\end{figure}

\begin{figure}[ht]
\centering
\noindent\includegraphics[width=3in,height=3in]{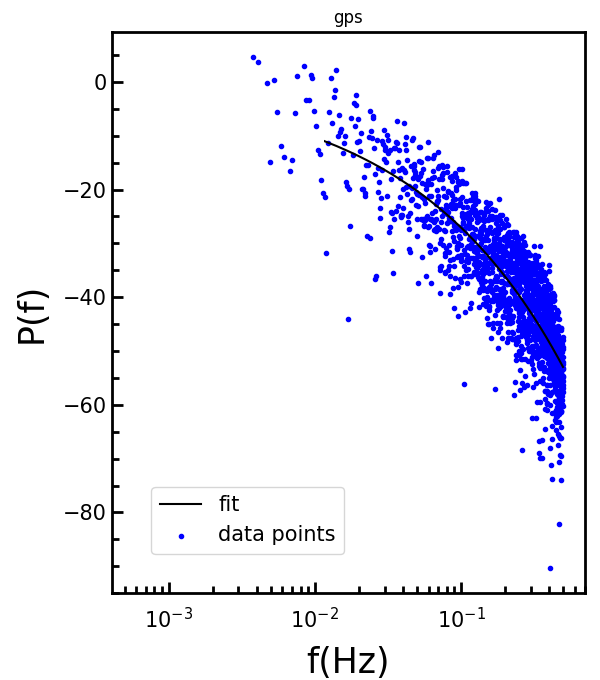}
\caption{The PSD variations with least square fit (black solid line) corresponding to the time bins of intense C/N$_o$ variation of Figure \ref{sc4-gps}.} 
\label{sc5-gps}
\end{figure}

The temporal power spectrum has been introduced by several researchers \cite{sc:14,sc:15,sc:16,sc:17} as follows: 
\begin{linenomath*}
\begin{equation}
S = \frac{T}{(f_0^2 + f^2)^\frac{p}{2}}
\label{sca}
\end{equation}
\end{linenomath*}
where f$_0$ is the outer scale frequency, T is the spectral strength and p is the spectral slope. Equation \ref{sca} can be simplified to the following (when $f >> f_0$):
\begin{linenomath*}
\begin{equation}
S = T f^{-p}
\label{scb}
\end{equation}
\end{linenomath*}
Furthermore, the spectrum of electron density fluctuations ($\delta N$) can be modeled as a power law, with an outer scale, given by \cite{sc:15,sc:16,sc:22}:
\begin{linenomath*}
\begin{equation}
S_{\delta N}(q) =  \frac{C_s}{(q_0^2+q^2)^{(m+\frac{1}{2})}}
\label{scc}
\end{equation}
\end{linenomath*}
where, $C_s$ is the strength of the turbulence proportional to T \cite{sc:15,sc:16,sc:22}, m is the irregularity spectral index (p=2m-1) and $q_0$ is the wave number of the outer scale and is related to the outer scale turbulence (L) as:
\begin{linenomath*}
\begin{equation}
L = \frac{2\pi}{q_0}
\label{scd}
\end{equation}
\end{linenomath*}
Therefore, for $q>>q_0$, the spectrum in equation \ref{scc} modifies to:
\begin{linenomath*}
\begin{equation}
S_{\delta N}(q) = C_sq^{-(2m+1)} = C_sq^{-(p+2)}
\label{sce}
\end{equation}
\end{linenomath*}
Figure \ref{sc3} shows the PSD variation from PRN5 and PRN6. The absolute values of p for the PRNs 5 and 6 are 3.690$\pm$0.009 and 3.596$\pm$0.008 respectively. These values are obtained from the co-variance matrix of the least squares fit performed to the PSD in the logarithmic domain by using equation \ref{scb}. These values are then verified with the approximate estimation \cite{sc:15,sc:16} of $p = \frac{[P(0.05)-P(0.5)]}{10}$ from figure \ref{sc3}. The values of the corresponding irregularity scale sizes (using equations \ref{scc} through \ref{sce}) are found out to be ranging from about 518 m to 6.026 km.    
The GPS observations of C/N$_o$ variance (top panel), standard deviation of $\frac{(C/N_o)}{<C/N_o>}$ (bottom panel) and the corresponding PSD are shown in Figures \ref{sc4-gps} and \ref{sc5-gps} respectively. In Figure \ref{sc4-gps}, it can be clearly observed that the C/N$_o$ variance was the most significant during the time bin of 16-17 UT. 
Subsequently, the corresponding PSD of that time bin has been plotted in Figure \ref{sc5-gps}. The value of p is 3.318$\pm$0.011 and the irregularity scale sizes are ranging from about 503 m to 6.007 km as observed by the GPS satellites on that day. It is to be noted that these observations are comparable to that observed by NavIC and hence the suitability of NavIC to detect and measure the ionospheric irregularities is quite evident. 

\begin{figure}[ht]
\centering
\noindent\includegraphics[width=5.5in,height=4in]{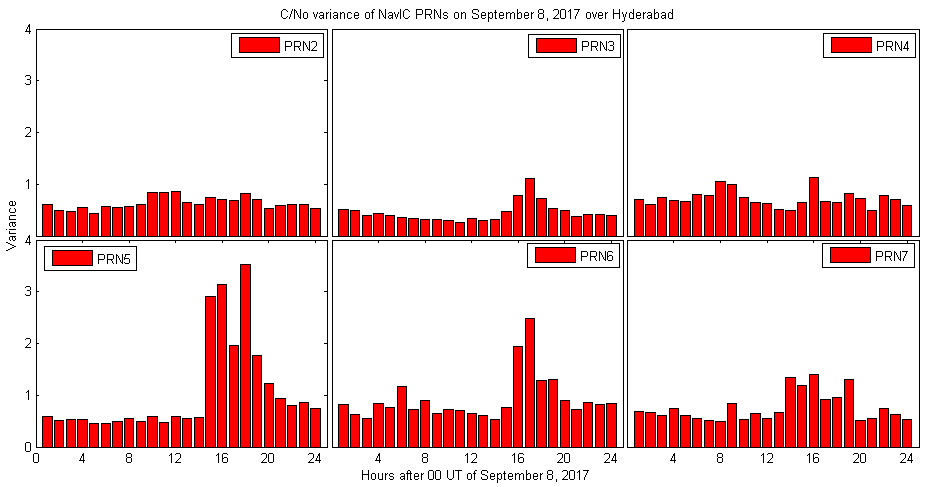}
\noindent\includegraphics[width=5.5in,height=4in]{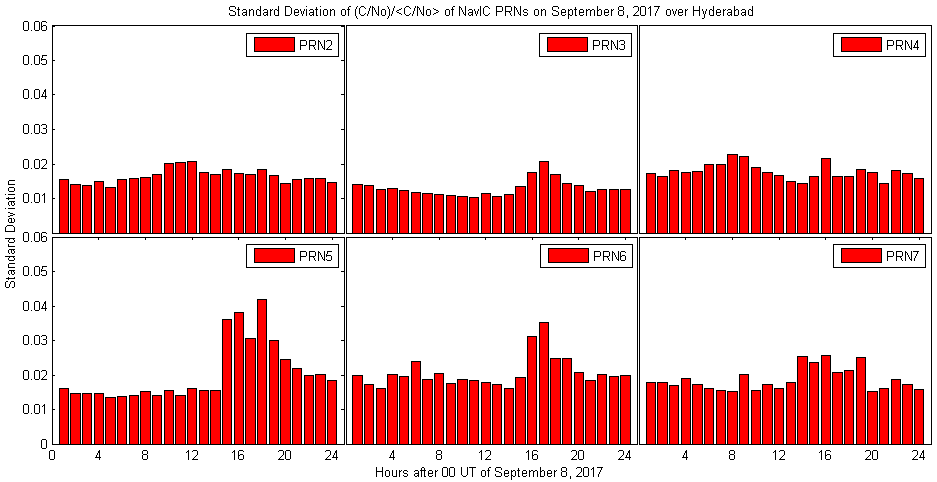}
\caption{The hourly binned variance plots (top panel) of C/N$_o$ and the corresponding standard deviation plots (bottom panel) of $\frac{(C/N_o)}{<C/N_o>}$ for all PRNs of NavIC on September 8, 2017 over Hyderabad.} 
\label{sc4-ou}
\end{figure}

\begin{figure}[ht]
\centering
\noindent\includegraphics[width=3in,height=2.5in]{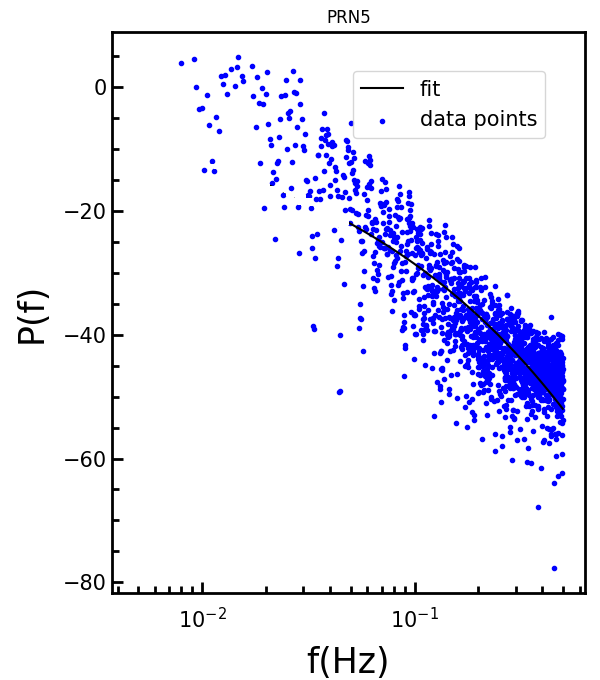}
\noindent\includegraphics[width=3in,height=2.5in]{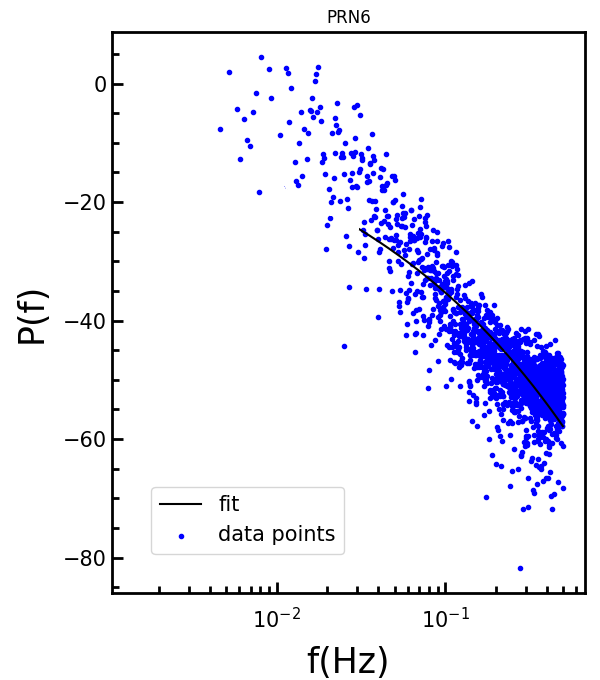}
\caption{The PSD variations with least square fit (black solid line) corresponding to the time bins of intense C/N$_o$ variation of Figure \ref{sc4-ou}.} 
\label{sc5-ou}
\end{figure}

Taking similar approach for the storm day of September 8, 2017, the following Figure \ref{sc3-ou} shows the NavIC measured C/N$_o$ variation (dB-Hz) over Hyderabad located away from the northern crest of the EIA (or closer to the magnetic equator) of the Indian region. 
Upon closely observing panels showing PRNs 5 and 6, it is clearly visible that there had been strong fluctuations in the C/N$_o$ with values dropping from $\sim$51 dB-Hz to 36 dB-Hz around 17-18 UT (22:30-23:30 LT) for PRN 5 and from $\sim$52 dB-Hz to $\sim$37 dB-Hz around 16-17 UT (21:30-22:30 LT), between local post-sunset and pre-midnight sector.
\begin{figure}[ht]
\centering
\noindent\includegraphics[width=5.5in,height=4in]{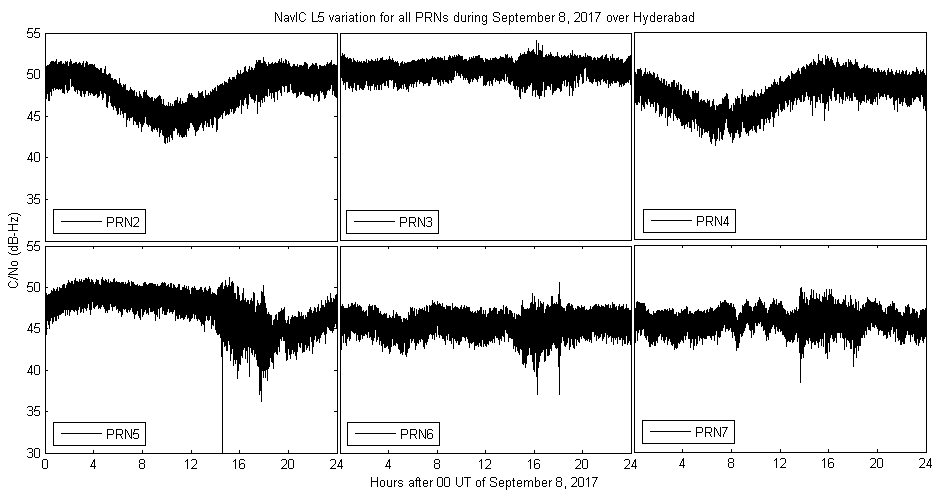}
\caption{The C/N$_o$ (dB-Hz) variation during the disturbed day of September 08, 2017, as observed by L5 signal of NavIC satellite PRNs 2-7 over Hyderabad.}
\label{sc3-ou}
\end{figure} 
Similarly, for verification of the observed C/N$_o$ drops due to scintillation, Figure \ref{sc4-ou} shows the hourly binned variance plot (top panel) and the corresponding standard deviation of $\frac{(C/N_o)}{<C/N_o>}$ plot (bottom panel) for all the PRNs of NavIC during the entire day of September 8, 2017. In the hourly bin of 17-18 UT for PRN 5 and 16-17 UT of PRN 6, it is clearly visible that the variance over the average values are significant.
Figure \ref{sc5-ou} shows shows the PSD variation from the PRNs 5 and 6. The values of p for PRNs 5 and 6 are 3.072$\pm$0.011 and 3.008$\pm$0.010 and the irregularity scale sizes are ranging from about 506 m to 6.039 km.

\clearpage
\subsection{September 16, 2017: a case of quiet condition}

\begin{figure}[ht]
\centering
\noindent\includegraphics[width=5.5in,height=4in]{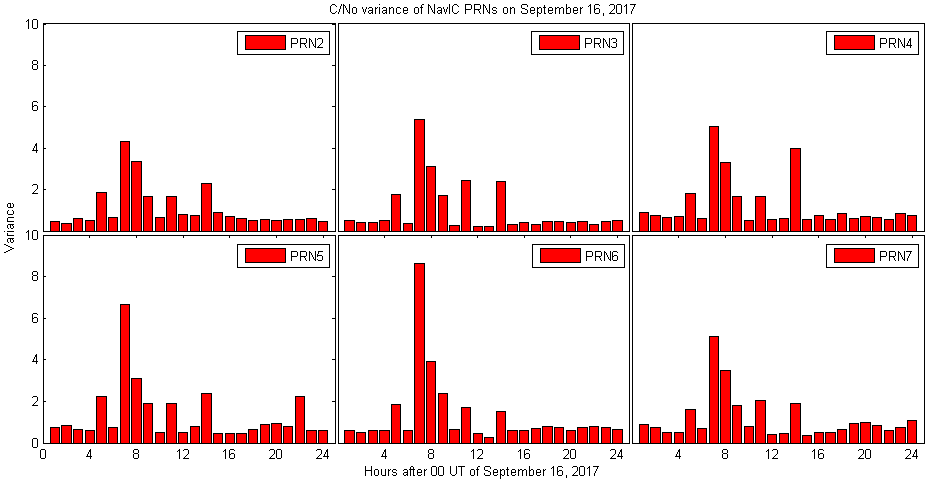}
\noindent\includegraphics[width=5.5in,height=4in]{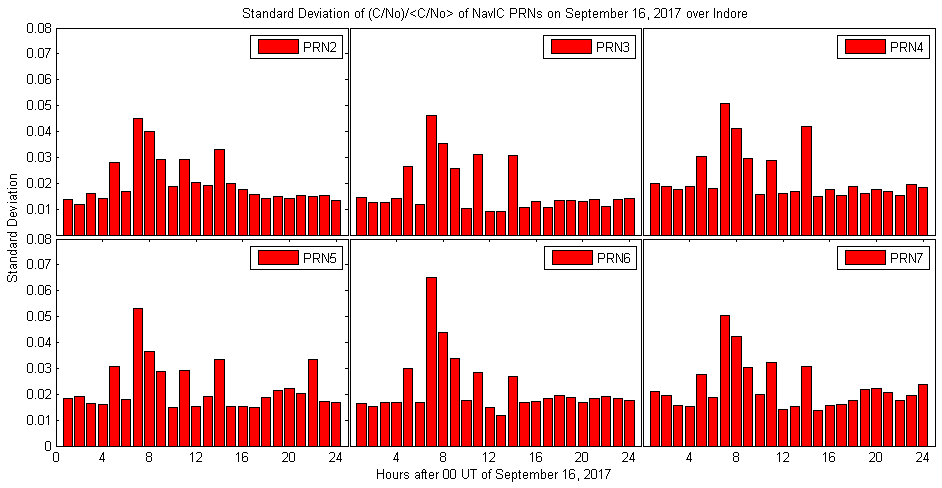}
\caption{The hourly binned variance plots (top panel) of C/N$_o$ and the corresponding standard deviation plots (bottom panel) of $\frac{(C/N_o)}{<C/N_o>}$ for all PRNs of NavIC on September 16, 2017 over Indore.} 
\label{sc7}
\end{figure}

\begin{figure}[ht]
\centering
\noindent\includegraphics[width=3in,height=2.5in]{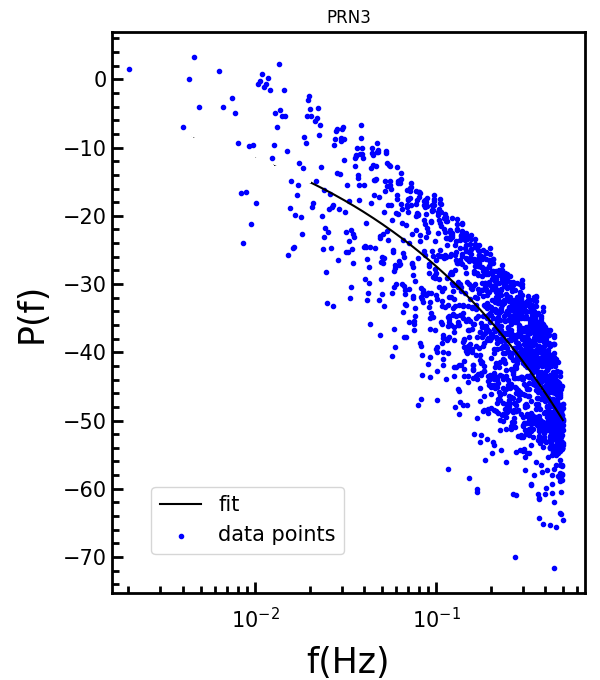}
\noindent\includegraphics[width=3in,height=2.5in]{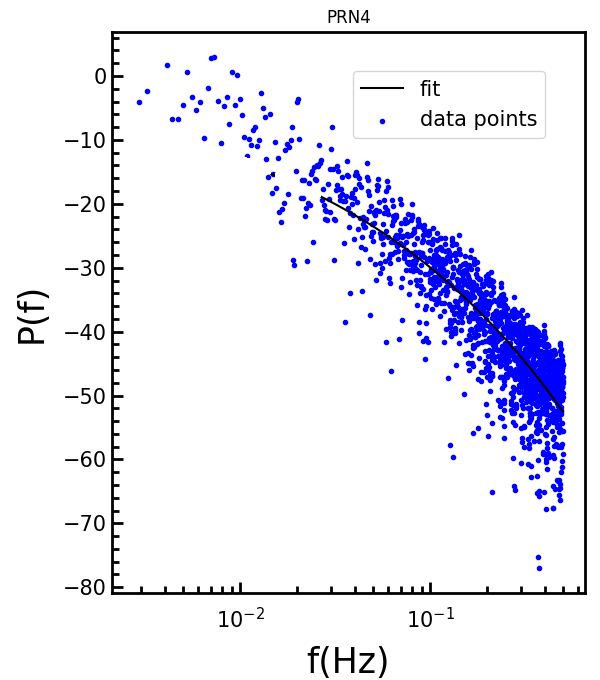}
\noindent\includegraphics[width=3in,height=2.5in]{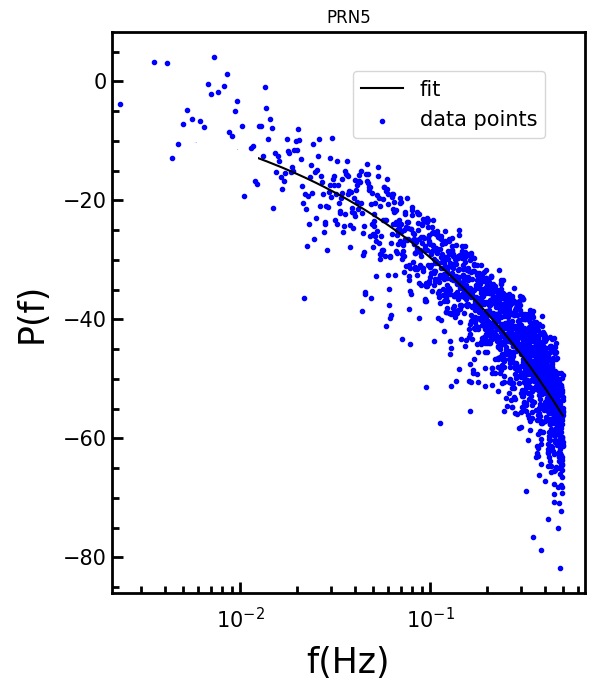}
\caption{The PSD variations with least square fit (black solid line) corresponding to the time bins of C/N$_o$ variation of Figure \ref{sc7}.} 
\label{sc8}
\end{figure}

Following similar approach as presented for the previous case, the following Figure \ref{sc6} shows the C/N$_o$ variation (dB-Hz) for the entire day of September 16, 2017, as observed by the L5 signal of NavIC. Drops in the C/N$_o$ has been observed at multiple time stamps throughout the day by all the PRNs. 
\begin{figure}[ht]
\centering
\noindent\includegraphics[width=5.5in,height=4in]{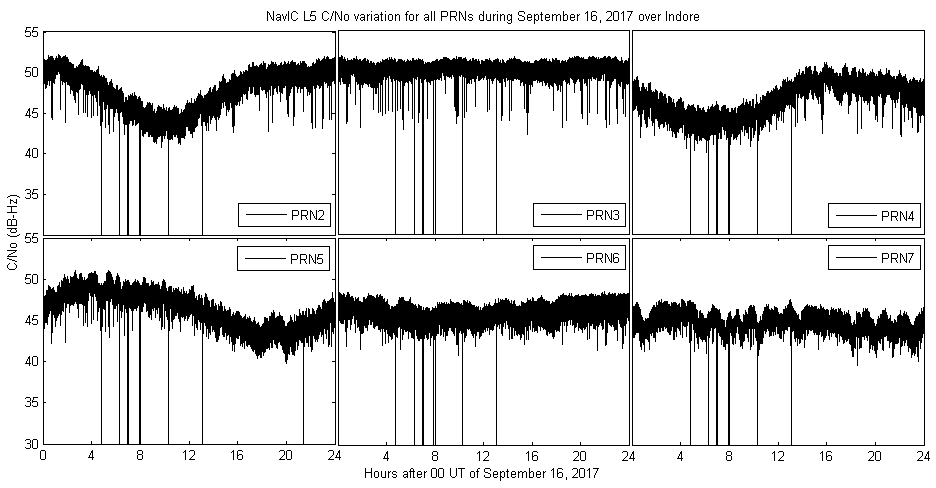}
\caption{The C/N$_o$ (dB-Hz) variation during the entire quiet day of September 16, 2017 over Indore, as observed by L5 signal of NavIC satellite PRNs 2-7.} 
\label{sc6}
\end{figure}
However, to confirm whether the C/N$_o$ drops had been significant, Figure \ref{sc7} has been plotted that shows the hourly binned variance plots (top panel) and the corresponding standard deviation plots (bottom panel) of $\frac{(C/N_o)}{<C/N_o>}$ of all the PRNs of NavIC during the entire day of September 16, 2017. The hourly bin of 6-7 UT for all the PRNs show the most significant rise and hence maximum variation among all the bins of the day. 
Figure \ref{sc8} shows the PSD variation from all the satellites of NavIC. It is to be noted that only PRNs 3, 4, and 5 show a power law variation in the PSD and hence are taken in consideration. The values of p for PRNs 3, 4 and 5 are 2.873$\pm$0.012, 2.949$\pm$0.009 and 2.829$\pm$0.008 respectively. The values of the corresponding irregularity scale sizes are ranging from about 500 m to 6.057 km.  
\clearpage

\subsection{August 17, 2018: a case of weak storm}

Due to the passage of a HSSW stream around the Earth, from a CH on the Sun, a G1 (K$_p$= 5, minor) level geomagnetic storm was observed on August 17, 2018. The Dst index showed dip on this day with a value of -37 nT at 05:00 UT, thereafter signifying this storm to be of a weak nature.

\begin{figure}[ht]
\noindent\includegraphics[width=5.5in,height=4in]{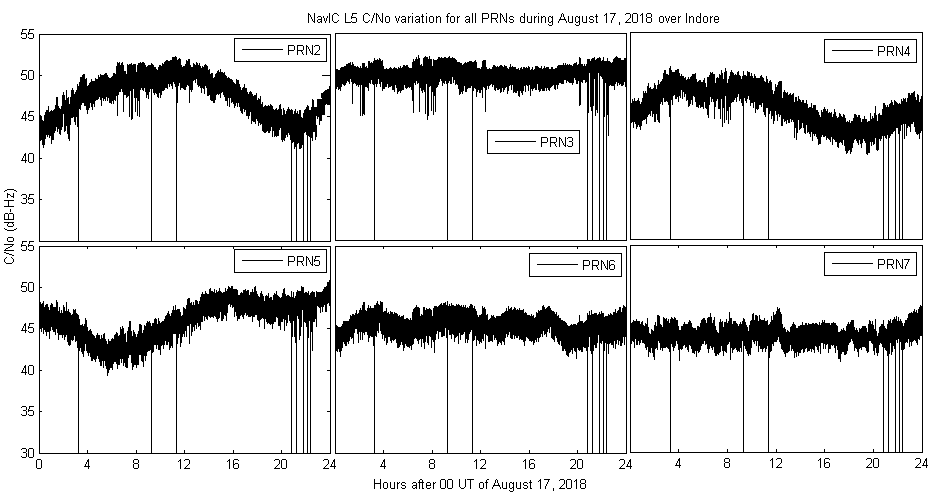}
\caption{The C/N$_o$ (dB-Hz) variation during the entire day of August 17, 2018 over Indore, as observed by L5 signal of NavIC satellite PRNs 2-7.} 
\label{sc9}
\end{figure}

\begin{figure}[ht]
\centering
\noindent\includegraphics[width=5.5in,height=4in]{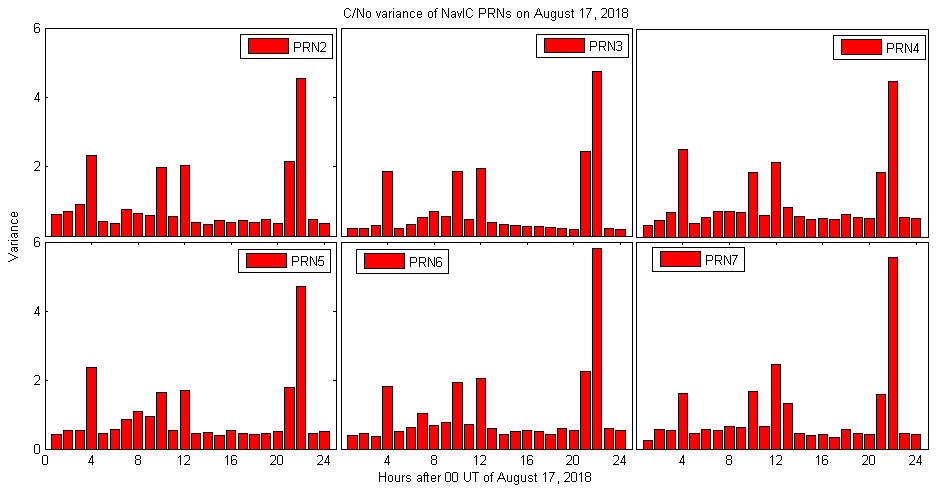}
\noindent\includegraphics[width=5.5in,height=4in]{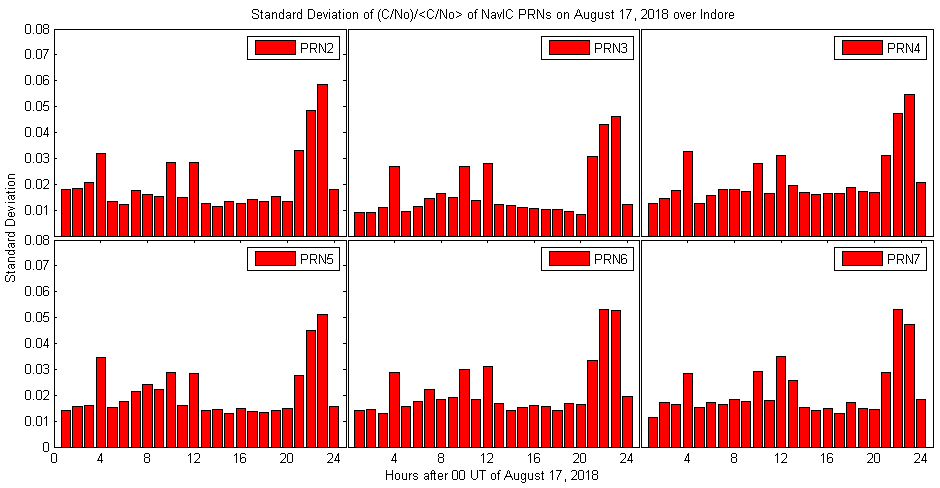}
\caption{The hourly binned variance plots (top panel) of C/N$_o$ and the corresponding standard deviation plots (bottom panel) of $\frac{(C/N_o)}{<C/N_o>}$ for all PRNs of NavIC on August 17, 2018 over Indore.} 
\label{sc10}
\end{figure}

\begin{figure}[ht]
\centering
\noindent\includegraphics[width=3in,height=2.5in]{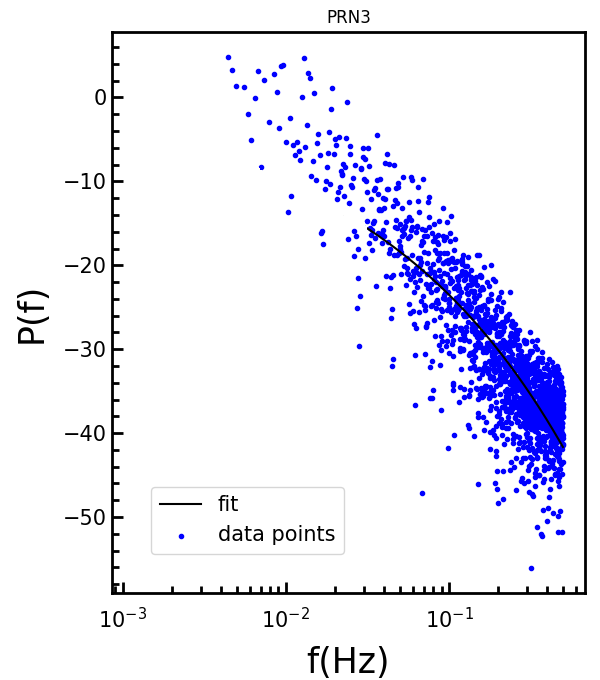}
\noindent\includegraphics[width=3in,height=2.5in]{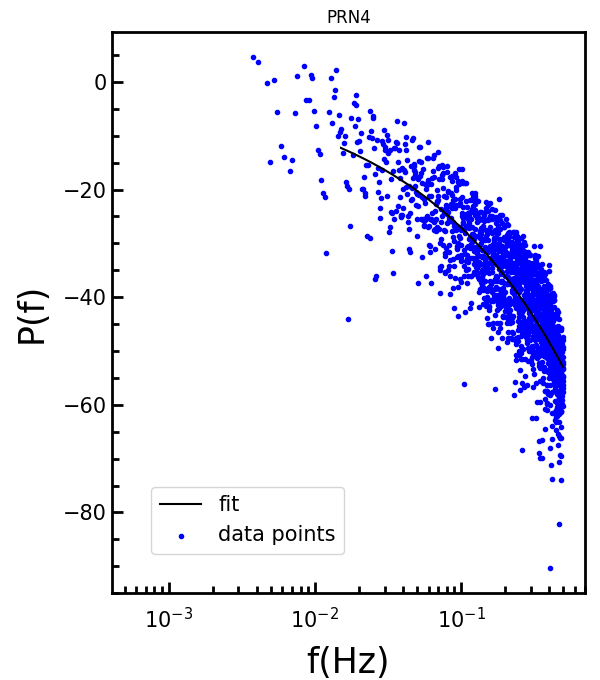}
\noindent\includegraphics[width=3in,height=2.5in]{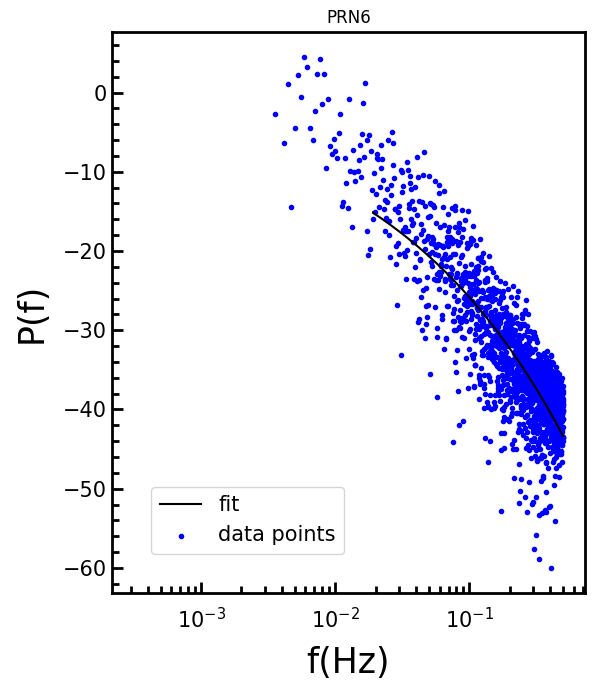}
\caption{The PSD variations with least square fit (black solid line) corresponding to the time bins of C/N$_o$ variation of Figure \ref{sc10}.}
\label{sc11}
\end{figure}

Very similar to the previous two cases, Figure \ref{sc9} shows the C/N$_o$ variation (dB-Hz) for the entire day of August 17, 2018, as observed by the L5 signal of NavIC. Drops in the C/N$_o$ has been observed at multiple time stamps throughout the day by all the PRNs.
However, for the verification of the significance of these C/N$_o$ drops, Figure \ref{sc10} shows the hourly binned variance plots (top panel) and the corresponding standard deviation plots (bottom panel) of $\frac{(C/N_o)}{<C/N_o>}$ of all the PRNs of NavIC during the entire day of August 17, 2018. The time bin of 21-22 UT for the all the PRNs show the most significant rise and hence maximum variation among all the bins of this day.
Figure \ref{sc11}, in a way similar to the previous cases, shows the PSD variation from all the NavIC satellites. It is to be noted that only PRNs 3, 4 and 6 show a power law variation in the PSD and hence are taken in consideration. The values of p for PRNs 3, 4 and 6 are 2.306$\pm$0.009, 2.518$\pm$0.011 and 2.322$\pm$0.008 respectively. The values of the corresponding irregularity scale sizes are ranging from about 497 m to 6.001 km, which is comparable to the previous cases.

Finally to understand the nature of the ionospheric variability and the corresponding irregularity scale sizes from simultaneous observations over locations near the northern crest of EIA and the magnetic equator, Tables \ref{tab1} and \ref{tab2} summarize the values of p and the range of the outer scale sizes (in km) corresponding to the NavIC satellite PRNs and the GPS satellites that showed power law variation from the temporal analysis, for the cases where there had been C/N$_o$ drops observed by the GPS L5 signals over Indore and NavIC L5 signals over both Indore and Hyderabad, under variable geophysical conditions and during the period: September 2017- September 2019 falling in the declining phase of solar cycle 24. The difference in the irregularity scale sizes observed by these these PRNs can be attributed to the different ionospheric pierce points corresponding these satellites.
\newpage
\begin{table}[ht]
\caption{Values of p and the corresponding range of outer scale size of irregularities for the cases when GPS and NavIC L5 C/N$_o$ drops were observed during varying geophysical conditions over Indore (a location near the northern crest of EIA).}
\centering
\begin{center}
\begin{tabular}{ccccc}
\hline
Day & Geophysical conditions & PRN & p & range of L (km) \\ \hline
Sep 08, 2017 & Strong storm condition & 5 & 3.690$\pm$0.009 & 0.518-6.026\\
                 & & 6 & 3.596$\pm$0.008 & 0.518-6.014 \\
                 & & GPS & 3.318$\pm$0.011 & 0.503-6.007 \\ \hline
Sep 16, 2017 & Quiet condition & 3 & 2.873$\pm$0.012 & 0.500-5.974 \\
                 & & 4 & 2.949$\pm$0.009 & 0.515-6.057  \\
                 & & 5 & 2.829$\pm$0.008 & 0.507-5.991  \\
                 & & GPS & 2.705$\pm$0.009 & 0.507-5.983  \\ \hline
Aug 17, 2018 & Weak storm condition & 3 & 2.306$\pm$0.009 & 0.499-5.993  \\
                 & & 4 & 2.518$\pm$0.011 & 0.497-5.834  \\
                 & & 6 & 2.322$\pm$0.008 & 0.503-6.001  \\
                 & & GPS & 2.242$\pm$0.013 & 0.510-6.054 \\ \hline
\end{tabular}
\end{center}
\label{tab1}
\end{table}
\begin{table}[ht]
\caption{Values of p and the corresponding range of irregularity scale size for the cases NavIC L5 C/N$_o$ drops were observed along with power law during the varying geophysical conditions over Hyderabad (a location in between crest and the magnetic equator).}
\centering
\begin{center}
\begin{tabular}{ccccc}
\hline
Day & Geophysical conditions & PRN & p & range of L (km) \\ \hline
Sep 08, 2017 & Strong storm condition & 5 & 3.072$\pm$0.011 & 0.506-6.007 \\
                 & & 6 & 3.008$\pm$0.010 & 0.519-6.039 \\ \hline
Sep 16, 2017 & Quiet condition & 3 & 2.910$\pm$0.018 & 0.496-5.948 \\
                 & & 4 & 2.816$\pm$0.007 & 0.511-6.025  \\ \hline
Aug 17, 2018 & Weak storm condition & 5 & 2.509$\pm$0.009 & 0.509-5.992 \\ \hline
\end{tabular}
\end{center}
\label{tab2}
\end{table}
\clearpage

\section{Conclusions}

The ionosphere over the Indian longitude sector is highly dynamic and geosensitive due to the presence of the northern crest of EIA and the magnetic equator. An extensive study of the ionospheric irregularities, that have drastic impacts on the performance of the global navigational satellite systems in and around these locations under varied solar and geophysical conditions, is essential. The detailed study of these ionospheric irregularities brings forward important aspects to understand the ionospheric physics and related processes. In order to address this problem, this paper presents the simultaneous characterization of the low-latitude ionospheric irregularities over locations chosen such as to cover the zones of both the near the northern crest of EIA (Indore: 22.52$^\circ$N, 75.92$^\circ$E geographic and magnetic dip of 32.23$^\circ$N, with observations from NavIC and GPS L5 signals) and in between crest and the magnetic equator (Hyderabad: 17.41$^\circ$N, 78.55$^\circ$E geographic and magnetic dip of 21.69$^\circ$N, with observations from NavIC L5 signals) of the Indian subcontinent. This study utilizes the L5 signal C/N$_o$ variations to analyse the power spectrum and determine the spectral slope and range of the ionospheric irregularity scale sizes.The study has been performed during a span of September 2017- September 2019 (declining phase of solar cycle 24) that consisted of weak, strong geomagnetic storms and quiet-time condition. Observations show that the irregularity scale size ranges from about 500 m to 6 km. These results are consistent with the previously reported \cite{sc:27} irregularity scale sizes associated to L-band scintillations over regions surrounding the EIA. This study uses actual transionospheric signals rather than signals based on in situ measurements and/or phase screen model derived analysis. The observations also show no correlation between the scintillation observed and the geomagnetic activity. The observations are indeed consistent with the fact that scintillations are associated with the formation and evolution of EPBs in and around the EIA region. While the formation of plasma depletions can be either suppressed or stimulated by geomagnetic storms, once formed the turbulence and associated scintillation is determined by the local plasma properties. This study for the first time, shows the nature of the temporal PSD for ionospheric scintillation during varying solar and geophysical conditions, by measuring the intermediate irregularity scale sizes that generally occur in the low-latitude regions, utilizing simultaneous observations from NavIC and GPS over locations near the northern crest of the EIA and in between crest and the magnetic equator which ensures proper ionospheric irregularity characterization over the geosensitive Indian longitude sector.

\acknowledgments

SC acknowledges Space Applications Centre (SAC), Indian Space Research Organization for providing research fellowship under the project NGP-17. DA acknowledges the INSPIRE fellowship from the Department of Science and Technology, which she used to pursue her research. The authors acknowledge SAC, ISRO for providing the NavIC receivers to Department of Astronomy, Astrophysics and Space Engineering, IIT Indore under this project. Authors also acknowledge the Department of ECE, Osmania University for providing us with the NavIC data. Further acknowledgements go to World Data Center for Geomagnetism, Kyoto for the Dst index freely available at http://wdc.kugi.kyoto-u.ac.jp/kp/index.html and for the sunspot numbers available at http://www.sidc.be/silso/ (SILSO data/image, Royal Observatory of Belgium, Brussels). The authors acknowledge the reviewers for their suggestions in improving the quality of the manuscript.

\newpage
\bibliography{a4}

\end{document}